\documentclass[fleqn]{annalen}
\usepackage{graphics}
\begin{document}
\newcommand{\volume}{8}              
\newcommand{\xyear}{1999}            
\newcommand{\issue}{5}               
\newcommand{\recdate}{29 July 1999}  
\newcommand{\revdate}{xxxx}    
\newcommand{\revnum}{0}              
\newcommand{\accdate}{xxxxxx}    
\newcommand{\coeditor}{xxx}           
\newcommand{\firstpage}{1}         
\newcommand{\lastpage}{10}          
\setcounter{page}{\firstpage}        
\newcommand{\keywords}{metal-insulator transition, 
interactions, quantum chaos} 
\newcommand{\PACS}{71.30.+h, 72.15.Rn, 05.45.Mt}
\newcommand{\shorttitle}{D. L. Shepelyansky and P. H. Song, 
Quantum ergodicity for electrons in 2D} 
\title{Quantum ergodicity for electrons in two dimensions}
\author{Dima L.\ Shepelyansky$^{1}$, and Pil Hun Song$^{2,3}$} 
\newcommand{\address}
  {$^{1}$Laboratoire de Physique Quantique, 
  UMR 5626 du CNRS, Universit\'e Paul Sabatier,\\
  F-31062 Toulouse Cedex 4, France\\ 
  $^{2}$Center for Theoretical Physics, Seoul National University,\\
  Seoul 151-742, Korea\\
  $^{3}$ Max-Planck-Institut f{\"{u}}r Kernphysik, Postfach 103980,\\
  69029 Heidelberg, Germany}
\newcommand{\email}{\tt  dima@irsamc2.ups-tlse.fr} 
\maketitle
\begin{abstract}
  We study the effect of electron-electron interaction
  on a two dimensional (2D) disordered lattice. For the case of
  two electrons the analytical estimates are presented 
  showing a transition from localized to delocalized states
  in a way similar to the Anderson transition in 3D.
  The localized phase corresponds to large values of the
  parameter $r_s$, which is determined by the ratio of the Coulomb
  and Fermi energies. The numerical investigations of 
  the spectral statistics $P(s)$ in a system with up to
  $N_p=30$ spin polarized electrons show the transition
  from the Poisson to the Wigner-Dyson distribution
  at a total electron energy independent of $N_p$.
  The relation to experiments on the metal-insulator
  transition in 2D is also discussed.
\end{abstract}

\section{Introduction}
\label{introduction}

Since 1979  it became clear that
non-interacting electrons are always localized in a 2D
disordered potential \cite{and79}. At present this result
is firmly confirmed by different analytical and numerical
studies (see the reviews \cite{rmph,kramer}) according to which
all one-particle eigenstates are exponentially localized 
in the case of orthogonal or unitary symmetry.
Delocalization is possible only for the symplectic symmetry
which is however not realized in the absence of 
spin-orbit interaction \cite{rmph,kramer}.
Therefore, the discovery of metallic behavior in 2D high mobility
samples by Kravchenko {\it et. al.} \cite{krav94}
attracted a great interest of the solid-state community
and pushed forward the question about a role
of electron-electron interactions in the localized phase.
Indeed, in many experiments of different groups 
\cite{krav94,krav96,popovic,canada,yael,alex,pudalov,ensslin,mills} 
the ratio of the Coulomb interaction energy $E_{ee}$ 
to the Fermi energy $\epsilon_F$ is characterized by a large 
dimensionless parameter
$r_s = 1/{\sqrt{\pi n_s} {a^*_B}} \simeq
E_{ee}/\epsilon_F$, where $n_s$ is the electron density in 2D,
and $a^*_B = \hbar^2 \epsilon_0/m^* e^2$, $m^*$, $\epsilon_0$
are the effective Bohr radius, electron mass and dielectric constant,
respectively. Typically in the above experiments the parameter
$r_s$ is ranged  in the interval 5 - 40 showing that these
experiments are well outside of the perturbative regime
corresponding to $r_s \ll 1$. Generally, in these experiments
the insulating increase of resistivity at low temperature $T$
is observed at large $r_s$ values (low charge density $n_s$),
while its metallic decrease at low $T$ is seen at lower $r_s$
(higher $n_s$). In addition, experimentally it was 
found that a sufficiently strong magnetic field parallel 
to the 2D plane destroys the metallic behavior
\cite{simonian,yoon}. These results show an important role
of the spin degrees of freedom.
However, we should note that the recent 
experiments \cite{okamoto} with n-SiGe samples demonstrate that
the metallic behavior persists even in very strong in plane magnetic
field when all spins are polarized.
Another important experimental
indication was found for a localized phase with a variable range hopping
(VRH) \cite{vrh1,vrh2}. Indeed, the experiments show that a prefactor
in the exponential VRH resistivity dependence on temperature
is phonon independent \cite{vrh2}. For the theoretical
explication of this result other physical mechanisms
of VRH should be found \cite{shklov} and it is possible
that their origin is related to the electron-electron 
interaction as it had beed discussed long ago in \cite{and78}.

From the theoretical view point the problem of interaction 
in the localized phase
is rather nontrivial. Indeed, due to localization
of noninteracting states, the
two interacting particles (TIP), with a short range
repulsive/attractive interaction and located on a distance
of one-particle localization length $l_1$ from each
other, always return to their center of mass
that enhances enormously their interaction. As a result
the two particles can propagate coherently on a distance
much larger $l_1$. This effect has been discussed
recently by different groups for a short range interaction
\cite{ds94,imry,pichard,oppen,jack,moriond} (see also the early
paper by Dorokhov for a strongly attractive case \cite{dor}).
However, in the experiments discussed above
the charge density is very low and the distance between
electrons (holes) is much larger than $a^*_B$ ($r_s \gg 1$).
In this situation the localized electrons interact
via the long range Coulomb interaction. Recently,
it has been shown \cite{ds99,moriond2} that in this situation
two interacting electrons can be delocalized 
even when they are separated by a distance $R \gg l_1$. Moreover, this
delocalization goes in a way very similar to the Anderson
transition in 3D. In this paper we discuss the physical
origin of this delocalization (Section II).
After that (Section III) we present the results
of extensive numerical studies of up to $N_p=30$
electrons (with polarized spins) on 2D disordered lattice.
In agreement with the first results presented in
\cite{song} we establish that the ground state remains localized
(nonergodic) but at moderate values of parameter $r_s$
the electrons become ergodic at very low total 
energy where their spectral statistics demonstrates
a transition from the Poisson to the Wigner-Dyson distribution.
The conclusion is presented in Section IV.

\section{Two interacting electrons in the 2D Anderson model}
\label{tip}

Let us consider a system of electrons with
polarized spins on a 2D disordered lattice (Anderson model).
Such polarization can be reached by a strong in plane magnetic
field but also one can speak about 
spinless fermions. The system is described by the Hamiltonian
\begin{equation}
H = V\sum_{<ij>} a^\dagger_i a_j + \sum_i w_i n_i
+ U \sum_{i>j} \frac{n_i n_j}{r_{ij}} .
\end{equation} 
Here $a^\dagger_i (a_i)$ is the fermion creation (annihilation) operator
at site $i$, the hopping between the nearest neighbors is $V$,
the diagonal energies $w_i$ are randomly distributed
within the interval $[-W/2,W/2]$ and $U$ is the strength of the
Coulomb interaction with $r_{ij}$ being the distance
between electrons at sites $i,j$.
In this notation $n_i=a^\dagger_i a_i$ is the
occupation number at site $i$. The electrons (particles)
are moving in a 2D cell of size $L \times L$ with 
periodic boundary conditions.
The Coulomb interaction is taken 
between electrons in one cell of size $L$ and with 8 charge images
in nearby 8 cells as in \cite{ds99}.  The number of particles $N_p$
and the cell size were varied within the intervals $2 \leq N_p \leq 30$  
and $8 \leq L \leq 31$.  With the notations of Eq.~(1), the
parameter $r_s$ is given by $r_s = U/(2V\sqrt{\pi\nu})$, 
where $\nu = N_p/L^2$ is the filling factor and $\epsilon_F = 4\pi \nu V$.  
The majority of our data have been 
obtained for $U/V =2$, $0.0048 < \nu < 0.03$ that corresponds to
$3.22 < r_s < 8.14$. When changing $N_p$ the filling
factor was kept approximately constant (nearest rational value)
by the appropriate choice of $L$.

For $U=0$ all states in the Hamiltonian (1) are localized
and the one-particle localization length
varies exponentially with $W$ in the limit of weak disorder:
$\ln l_1 \sim (V/W)^2$. The numerical diagonalization 
\cite{ds99,moriond2}
allows to determine the inverse participation ratio (IPR)
for one particle $\xi_1$, which approximately gives the number of 
lattice sites
contributing in one eigenstate. The values found at $L=24$
for different disorder strength at the ground state
and the middle of the band are correspondingly: 
$\xi_1$ = 3.4 and 4.2 $(W/V=15)$; 5.2 and 11.2 $(W/V=10)$; 
8.2 and 36.7 $(W/V=7)$; 13.5 and 84.2 $(W/V=5)$. This shows that 
the one-particle states at low energy and 
even at the band center (except maybe $W/V=5$)
are well localized $(\xi_1 \ll L^2)$.

To understand how two electrons interact in the 2D localized
phase we should estimate the interaction induced
transition rate $\Gamma_2$ between unperturbed $(U=0)$ localized
eigenstates. Following \cite{ds99,moriond2} let us assume
that the distance between electrons $R$
is much larger than $l_1$ ($R \gg l_1$) and their energy is in the 
middle of the band. In this case the two-body interaction
has a dipole-dipole form since the lower order terms
only slightly modify the one-particle effective potential.
The two-body matrix element between noninteracting
eigenstates is then $U_s \sim U l_1^2/R^3 \times \sum \psi^4$,
where we used that the dipole is of the order of $l_1$
(electron cannot jump on a distance larger $l_1$ due to exponential
localization). The sum $\sum $ of the product of four random
one-particle wave functions $\psi \sim 1/l_1$ runs over $l_1^2$ sites
for each electron so that $\sum \sim 1/l_1^2$ and $U_s \sim U/R^3$.
The density of directly coupled two electron states
is $\rho_2 \sim l_1^4/V$ and according to the Fermi golden rule
the hopping rate is $\Gamma_2 \sim U_s^2 \rho_2 \sim U^2 l_1^4/ (V R^6)$.
The mixing of two-electron levels takes place when
$\kappa_e = \chi_e^2 \sim \Gamma_2 \rho_2
\sim (r_L^{4/3}/r_s)^2 > 1$, where $r_L$ is the value of $r_s$
at the filling $\nu=1/l_1^2$ 
(one electron in a box of size  $l_1$ with $r_L=l_1 U/(2 \sqrt{\pi} V$)).
For $U \sim V$ this mixing takes place at $R \sim l_1^{4/3} \gg l_1$.
In this sense the situation is qualitatively different from the 
case of short range interaction where the mixing is possible
only for particles on a distance $R < l_1$. For $\kappa_e > 1$
the pair of two electrons becomes delocalized in a way similar to the
3D Anderson transition. Indeed, the center of mass can move
in the 2D plane and in addition electrons can rotate around a ring of
radius $R$ and width $l_1$ keeping their total energy
$E \sim U/R$ constant. Since $R >> l_1$ the dynamics is going
in an effective 3D space where the Anderson transition to diffusion
takes place at $\kappa_e > 1$. Formally, the diffusion of two electrons
will be finally localized since the finite length of the ring
makes the situation analogous to quasi-2D Anderson model
(finite number of planes). However, the 
corresponding TIP localization length $l_c$ jumps from $l_c \sim l_1$
at $\kappa_e < 1$ to exponentially
large $l_c \sim l_1 \exp(\pi l_1^{1/3} \kappa_e) \gg l_1$
for $\kappa_e > 1$. In a similar way for $M$ coupled 2D Anderson models
the one-particle localization length $l_1$ makes a jump near 
the 3D transition point (e.g. $W/V=16.5$)
from $l_1 \sim 1$ to $l_1 \sim \exp(g)$ where the conductance
$g \sim M$. In the delocalized phase the diffusion rate or the conductivity
in units $e^2/h$ can be estimated as $D_e \sim l_1^2 \Gamma_e \sim V
\kappa_e/l_1^2$. The above estimates show that 
the border between the insulating 
(localized at $r_s > r_L^{4/3}$) and metallic 
(delocalized at $r_s < r_L^{4/3}$) phases is:
\begin{equation}
r_s \sim r_L^{4/3}
\end{equation} 
The physical meaning of this result
is very natural: larger $r_s$ corresponds to a lower density $n_s$
with larger distance between electrons and smaller
two-body interaction between them.

The first numerical studies of two polarized electrons in 2D Anderson model
showed a transition in the level spacing  statistics $P(s)$
from the Poisson distribution $P_P(s)$ near the ground state to
the Wigner-Dyson statistics $P_W(s)$ at higher energies \cite{ds99}
in a way similar to the results of Shklovskii {\it et al.} \cite{shklovps}
for the 3D Anderson model. In both cases
the intermediate statistics $P(s)$ are very close 
at the critical transition point (Fig. 2 in \cite{ds99}).
The numerical studies were done in the range of parameters
$5 \leq W/V \leq 15$, $0.1 \leq U/V \leq 2$, $6 \leq L \leq 24$,
$ r_s < 10$; more details can be found in \cite{ds99,moriond2}.
Independent recent studies \cite{cuevas} of a similar model
with Coulomb interaction at the center of the energy band 
also show a transition to $P_W(s)$ as a function of $W$.
Since even analytical formulas
for the Coulomb matrix element between localized states are absent
further investigations of this problem are still needed.

\section{Numerical studies of multi-electron problem in 2D}
\label{mep}

For numerical studies of the Hamiltonian (1) we follow
the approach developed in \cite{ds99,song} and rewrite
the hamiltonian matrix in the basis of one particle
eigenstates (orbitals) at $U=0$ using the computed two-body
matrix elements between the orbitals. Only a finite
number $M$ of low energy orbitals has been considered
and the final many-body hamiltonian matrix 
was constructed on the basis of a pyramid rule
for the one-particle orbital index $m_i$: 
$\sum_{i=1}^{N_p} m_i \leq \sum_{i=1}^{N_p-1} i + M$.
This procedure allows to make an efficient reduction
of the resulting matrix size $N_m$ 
comparing to $_{M} C_{N_p}$ without any serious 
modification of the properties of low energy states.
To the maximum we used $N_m \approx 5000 $ corresponding
to $N_p =20, M=42$ (for $N_p=30$ we used up to $M=49$).
We checked that our results at low energy are not
sensitive to the variation of $M$ and $N_m$ (see
the illustrations below). One of the main characteristics
we extracted from the numerical studies is the
level spacing statistics $P(s)$ at a given
total excitation energy $E$ measured from the ground state.
The disorder average has been performed over
$N_D=5000$ realization (for low $E$) and
$N_D=1000$ (for higher $E$). In this way,
the total statistics used for $P(s)$
varied from $N_s=10^4$ at low $E$ to $N_S=3 \times 10^5$
at high $E$. To study the variation of $P(s)$
with the interaction or disorder strength 
it is convenient to use the parameter $\eta$ which is
defined as $\eta=\int_0^{s_0}
(P(s)-P_{W}(s)) ds / \int_0^{s_0} (P_{P}(s)-P_{W}(s)) ds$,
where  $s_0=0.4729...$ is the smaller intersection point of $P_P(s)$ and
$P_{W}(s)$.  In this way $\eta=1$ corresponds to $P_P(s)$ 
and $\eta$=0 to $P_{W}(s)$. We remind that for the
one-particle 3D Anderson model the localized phase
is characterized by $\eta=1$, since localized (nonergodic) eigenstates
do not feel each other, while in the metallic phase the eigenstates
are extended and $\eta=0$ \cite{shklovps}.

\begin{figure}
\centerline{\resizebox{9.44cm}{7.23cm}{\includegraphics{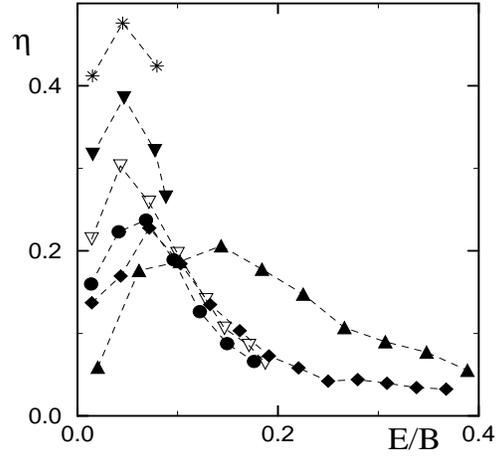}}}
\caption{Dependence of $\eta$ on the rescaled total energy $E/B$ for
various numbers of particles $N_p = 3$ (full
triangle up), 6 (full diamond),  10 ($\bullet$), 12 (triangle down), 
20 (full triangle down) and 30 (*); $W/V = 5$,  filling
factor $\nu \approx 1/32$ and $r_s = 3.22$, $10\leq L\leq 31$; $B=4 V$.
}
\label{fig1}
\end{figure}

The results presented in \cite{song} show that at $r_s=3.22$
the parameter $\eta$ evolves 
to 1 for the total energy $E<E_c$ and 0 for $E>E_c$
with the growth of the system size $L$ at fixed filling $\nu \approx 1/32$.
The critical energy $E_c$ and $\eta_c$ at which the transition
takes place are $E_c \approx 0.25 B $, $\eta_c \approx 0.56$ 
($ W/V=10$) and $E_c \approx 0.15 B $,  $\eta_c \approx 0.33$ ($ W/V=7$)
with $B=4 V$. For $W/V=15$ 
no transition is found for $E/B \leq 1$ where $\eta \approx 0.8 - 1.0$.
The new data for $W/V=5$ are presented in Fig. 1
where the transition is seen at $E_c \approx 0.1 B$, $\eta_c \approx 0.19$.
At the critical point $(E_c, \eta_c)$ the spectral statistics $P(s)$
is independent of the number of particles and the system size
reached in our simulations (see Fig. 3 in \cite{song}).
In Fig.2 we show an illustration that at low energy
the parameter $\eta$ is not sensitive to the variation of the total matrix
size $N_m$ by more than 3 times that confirms the validity
of our numerical approach. We note that not only the values
of $\eta$ are close for different $N_m$ but also the whole distributions
$P(s)$, as it is illustrated in Fig.3 for $r_s=5.76$. 

The results of Fig. 1 are obtained at fixed parameter $r_s$. 
The variation of the spectral statistics $P(s)$ 
at low energy for the change of $r_s$ from $3.2$ to $8.14$
is shown in Fig. 3. These data demonstrate a clear 
approach to the Poisson distribution with growing $r_s$.
The same tendency is seen in the $\eta$ dependence
on energy shown at different $r_s$ in Fig. 4. 
These results are in agreement with the relation (2)
and the argument given in the  previous section
according to which at large $r_s$ 
the distance between electrons grows and the two-body
interaction between them drops. Since 
\begin{figure}
\centerline{\resizebox{9.44cm}{7.23cm}{\includegraphics{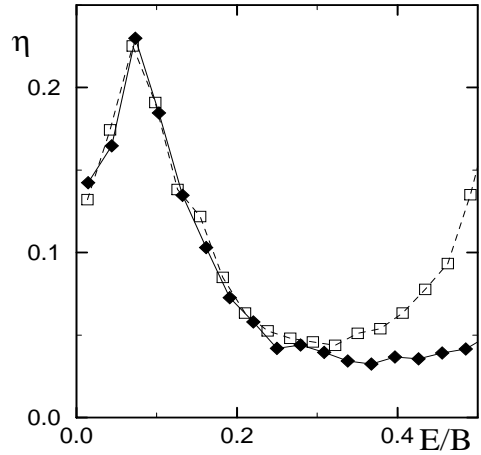}}}
\caption{Dependence of $\eta$ on $E/B$ for the case $N_p=6$ of Fig.1
for different matrix sizes: $N_m=1513, M=26$ (full diamond);
$N_m=498, M=21$ (square); here $N_D=5000$.
}
\label{fig2}
\end{figure}

\begin{figure}
\centerline{\resizebox{9.44cm}{7.23cm}{\includegraphics{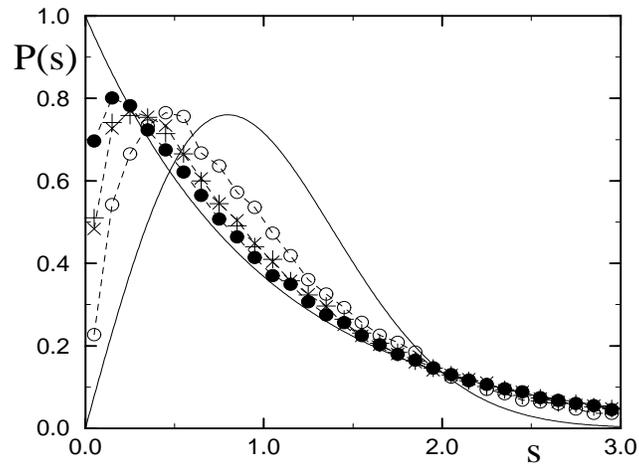}}}
\caption{Level statistics $P(s)$ at $W/V=7$, $L=25$
in the energy interval $0.1 < E/B < 0.14$ for different $r_s$:
$r_s=8.14$, $N_p=3$, $\eta=0.88$ ($\bullet$);
$r_s=5.76$, $N_p=6$, $\eta=0.77, M=27, N_m=1844$ ($\times$) (compare with
(+) at smaller basis $M=24, N_m=996, \eta=0.78$);
$r_s=3.15$, $N_p=20$, $\eta=0.52$ (o);
total statistics $N_S > 2.5 \times 10^4$. Full lines show the Poisson
distribution and the Wigner surmise.
}
\label{fig3}
\end{figure}
only the interaction between two particles
is able to mix many-body levels,
the electron dynamics becomes nonergodic at large $r_s$
that leads to the Poisson distribution. 

At the same time
even at the optimal values of $r_s \approx 3$ the ergodicity and
level mixing appear only at total energy $E>E_c$
while for $E<E_c$ including the ground state the ergodicity
is absent in the range of parameters we studied. Formally,
one can object against the statement that the value $\eta=1$
implies  the localization of electrons in space
giving as a counterexample  the case of
noninteracting delocalized electrons
in the metallic phase where also $\eta=1$ for the multi-electron spectrum.
However, in the system we study  all one-particle states
are exponentially localized and it is very difficult to imagine
that the interaction gives delocalization of charge,
makes the system metallic and does not introduce
complete ergodicity and level mixing, which should give
$\eta=0$ instead of $\eta=1$ found for $E<E_c$. Due to this
reason we consider that the appearance of
Poisson statistics near the ground state is the direct 
evidence of localization at $E<E_c$. At the same time
the emergence of random matrix statistics at $E>E_c$
implies ergodicity and delocalization for many electron
states. This delocalization apparently appears
in a way similar to the two electron delocalization discussed
in the previous section but in addition the
interaction between larger number
of particles decreases the delocalization border comparing to the TIP case
discussed in \cite{ds99,moriond2}.
An interesting point is that the transition at $E=E_c$
takes place at zero temperature. Indeed, according to the data
of Fig. 1 at the transition the energy per particle 
$\epsilon_c=E_c/N_p \approx 5 \times 10^{-3} B$
is much smaller than the Fermi energy $\epsilon_F = 0.1 B$
showing that the data are close to the thermodynamic limit
with $T=0$.

\begin{figure}
\centerline{\resizebox{9.44cm}{7.23cm}{\includegraphics{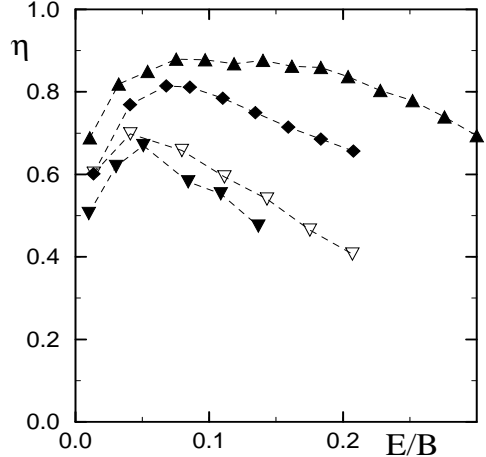}}}
\caption{Dependence of $\eta$ on $E/B$ for
different $r_s$ at $L=25$:
$r_s=8.14$, $N_p=3$ (full
triangle up); $r_s=5.76$,  $N_p=6$ (full diamond);
$r_s=4.07$, $N_p= 12$ (triangle down);
$r_s=3.15$, $N_p=20$ (full triangle down).
}
\label{fig4}
\end{figure}

\section{Conclusion}

Our results show that the Coulomb interaction between
two electrons in excited states leads to their delocalization
for $1< r_s < r_L^{4/3}$ while for $r_s > r_L^{4/3}$
they remain localized. The transition between these phases
is similar to the Anderson transition in an effective dimension
$d_{eff}=3$.
The numerical studies of the
spectral statistics for many 
polarized electrons in the 2D Anderson model 
show that for $3 < r_s < 9$ and $5 \leq W/V \leq 15$
the ground state is nonergodic (localized)
and is characterized by the Poisson statistics
for the total energy $E< E_c$. However,
the transition to quantum ergodicity
and the Wigner-Dyson statistics takes place
at a fixed total energy $E_c$ independent
of system size (for $r_s \approx 3.2$,
$5 \leq W/V \leq 10$ and fixed filling $\nu \approx 1/32$).
This implies a delocalization
at zero temperature $T$. At the critical point $E_c$ 
the parameter $\eta_c$ increases with the disorder
strength $W$: $\eta_c=0.19 \; (W/V=5); \; 0.33 \; (W/V=7);
\; 0.56 \; (W/V=10)$ and the critical statistics approaches
to the Poisson limit. In a certain sence the situation
is similar to the Anderson transition in 
high dimensions $d > 3$ discussed in \cite{isa}
where also $\eta_c$ is growing with $d$.
In analogy with this result and the case
of two electrons in 2D we make a conjecture
that in the Hamiltonian (1) the transition
at $E_c$ is similar to a transition in some
effective dimension $3 \leq d_{eff} < 2 N_p$.
This $d_{eff}$ is growing with the disorder strength $W$.

The interaction induced
ergodicity at $T=0$ is in favor of the metal-insulator
transition observed experimentally, especially in the view
of recently observed metallic behavior for 
polarized electrons \cite{okamoto}.
However, our data are not sufficient to determine the 
behavior of resistivity on temperature in the ergodic phase
at $E>E_c$. Therefore, it is not excluded
that at strong disorder this ergodic phase will
show a resistivity growth at low $T$. In this case
one can suppose that the ergodicity induced 
by the Coulomb interaction
is responsible for the phononless VRH conductivity as
it was argued in \cite{and78,shklov} and indicated
by the experiments \cite{vrh1,vrh2}.
More detailed investigations are required to understand
the properties of the Coulomb ergodic phase
at $E>E_c$.

\vspace*{0.25cm} \baselineskip=10pt{\small \noindent We acknowledge 
the IDRIS at Orsay for the allocation of the CPU time on supercomputers.
}
%
%
%
%
%
%
%
%
%
%
%
%


\begin{thebibliography}{9}
\bibitem{cuevas} E.~Cuevas, Phys. Rev. Lett. {\bf 83} (1999) 140;
                 we note that for the short range interaction
                 the majority of states is localized \cite{ds94,imry}
                 so that contrary to the claim of the paper
                 $P(s) \rightarrow P_P(s)$ for large $L$ at any 
                 $W \neq 0$.
\end{thebibliography}
\end{document}